\documentclass[fleqn,12pt,twoside]{article}
\usepackage[headings]{espcrc1}

\readRCS
$Id: espcrc1.tex,v 1.2 2004/02/24 11:22:11 spepping Exp $
\usepackage{graphicx}
\usepackage[figuresright]{rotating}
\usepackage[figuresright]{rotating}

\newcommand{\AmS}{{\protect\the\textfont2
  A\kern-.1667em\lower.5ex\hbox{M}\kern-.125emS}}

\title{Study of $J/\psi$ Production in $\sqrt{s_{NN}} = 200$ GeV $p+p$ and $d+Au$ Collisions in PHENIX}

\author{W. Xie\address{Riken-BNL Research Center, Brookhaven National Lab, Upton NY, 11973, USA} for the PHENIX Collaboration\thanks{for the full list of PHENIX authors and acknowledgements, see Appendix 'Collaboration' of this volume.}}

\begin{document}

\maketitle

\begin{abstract}
$J/\psi$ measurements in $p+p$ and $d+Au$ collisions serve as crucial 
references to understand the $J/\psi$ production in $Au+Au$ collisions
 at RHIC where quark gluon plasma (QGP) is expected to be formed.  They also 
provide important clues to study various interesting phenomena 
such as the gluon shadowing and color glass condensate.  We report the latest 
results from PHENIX experiment on $J/\psi$ production 
in $p+p$ and $d+Au$ collisions at forward, backward and midrapidity.
 
\end{abstract}

\section{Introduction}
$J/\psi$ enhancement{\cite{reconb} or strong suppression\cite{Satz86} is 
considered to be the signature of QGP formation. To quantify these effects, 
one needs measurements in $p+p$ and $d+Au$ collisions to 
understand many cold nuclear effects such as gluon shadowing, nuclear 
absorption,  initial state energy loss, Cronin effect and gluon 
saturation\cite{ppg038} that can have interesting impact on the $J/\psi$ 
production.

PHENIX\cite{PHENIX} measures $J/\psi$ via dielectron channel using midrapidity ($|\eta|<0.35$) central arm spectrometer and dimuon channel using forward ($1.2 < \eta < 2.4$) and backward ($-2.2 < \eta < -1.2$) rapidity muon arm spectrometer.

In year 2003, RHIC delieved 2.74~nb$^{-1}$ $d+Au$ and 350~nb$^{-1}$ $p+p$ 
collisions.  PHENIX central (muon) arm collected about 400 (1400) $J/\psi$ in
$d+Au$ and 100 (420) $J/\psi$ in $p+p$ collisions. These are the first measurements of cold nuclear effect at RHIC\cite{ppg038}. The final results are reported here.

\section{Results and discussion}

Figure \ref{fig:rapidity}(a) shows the $J/\psi$ rapidity distribution in $p+p$ 
collisions. The dashed error bars represent systematic uncertainties relevant for comparing the two rapidity bins in each muon arm, while the solid error bars represent the overall uncertainties relevant for comparing points at different rapidity. A $10\%$ of overall normalization error is not included. The total cross 
section is calculated via fitting the data with the PYTHIA predicted rapidity 
shape using GRV94HO parton distribution function (PDF) and the $J/\psi$ 
dilepton decay branch ratio of $5.9\%$.  The uncertainty from the PDFs is found to be less than $3\%$ after repeating the calculation using PDFs with very 
different shapes. The $\sigma_{pp}^{J/\psi} = 2.61 \pm 0.20 ({\rm fit}) \pm 0.26 ({\rm abs}) ~\mu$b, where the first error is the fitting error and the second one comes from the uncertainty of our minimum-bias trigger
bias. The result is consistent 
with our measurements in year 2002 \cite{ppg17}. Figure \ref{fig:rapidity}(b) 
shows the minimum-bias $R_{dAu} = \sigma_{dAu}/(2\times197\times\sigma_{pp})$ 
versus rapidity ($12\%$ overall normalization error is not included).  The forward (backward) rapidity corresponds to Bjorken variable $x_{bj} \sim 0.003$ ($0.01$),  where one expected shadowing (antishadowing) to be effective.  The midrapidity corresponds to $x_{bj} \sim 0.01$ where minimum 
nuclear effects are expected.  The EKS98 prediction\cite{Vogt2004} 
assuming 1mb 
absorption cross section fits the data best. The models assuming large 
shadowing effects like FGS \cite{Vogt2004} and coherence-length model \cite{KopelivichCoherence} are disfavored. 

\begin{figure}[htb]
\begin{minipage}[t]{81mm}
\includegraphics[width=21pc, height=15pc]{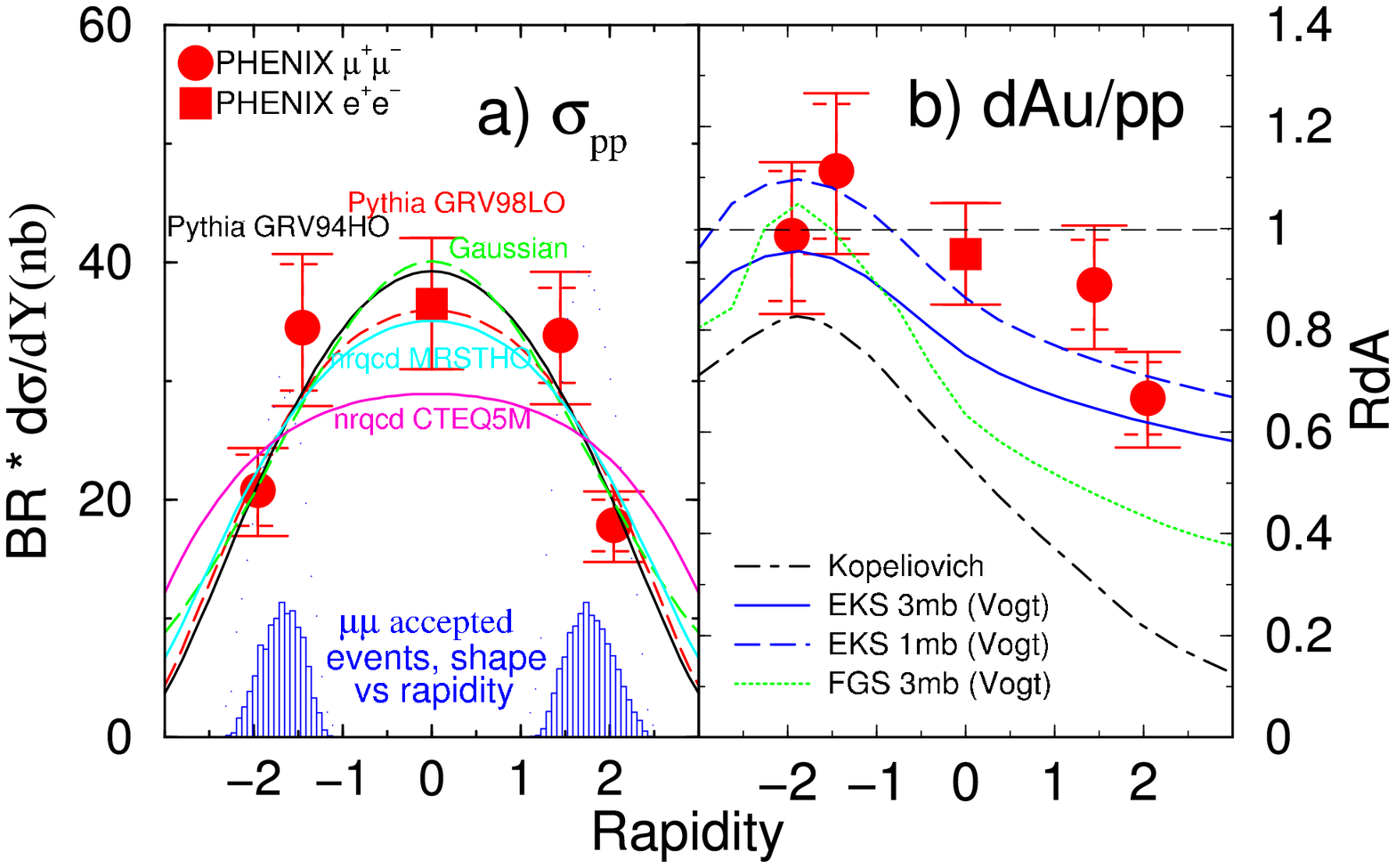}
\caption{\label{fig:rapidity} (a): Rapidity distribution in $p+p$ collisions. (b): The minimum bias $R_{dA}$ versus rapidity.} 
\label{fig:rapidity}
\end{minipage}
\hspace{\fill}
\begin{minipage}[t]{65mm}
\includegraphics[width=14pc, height=14.5pc]{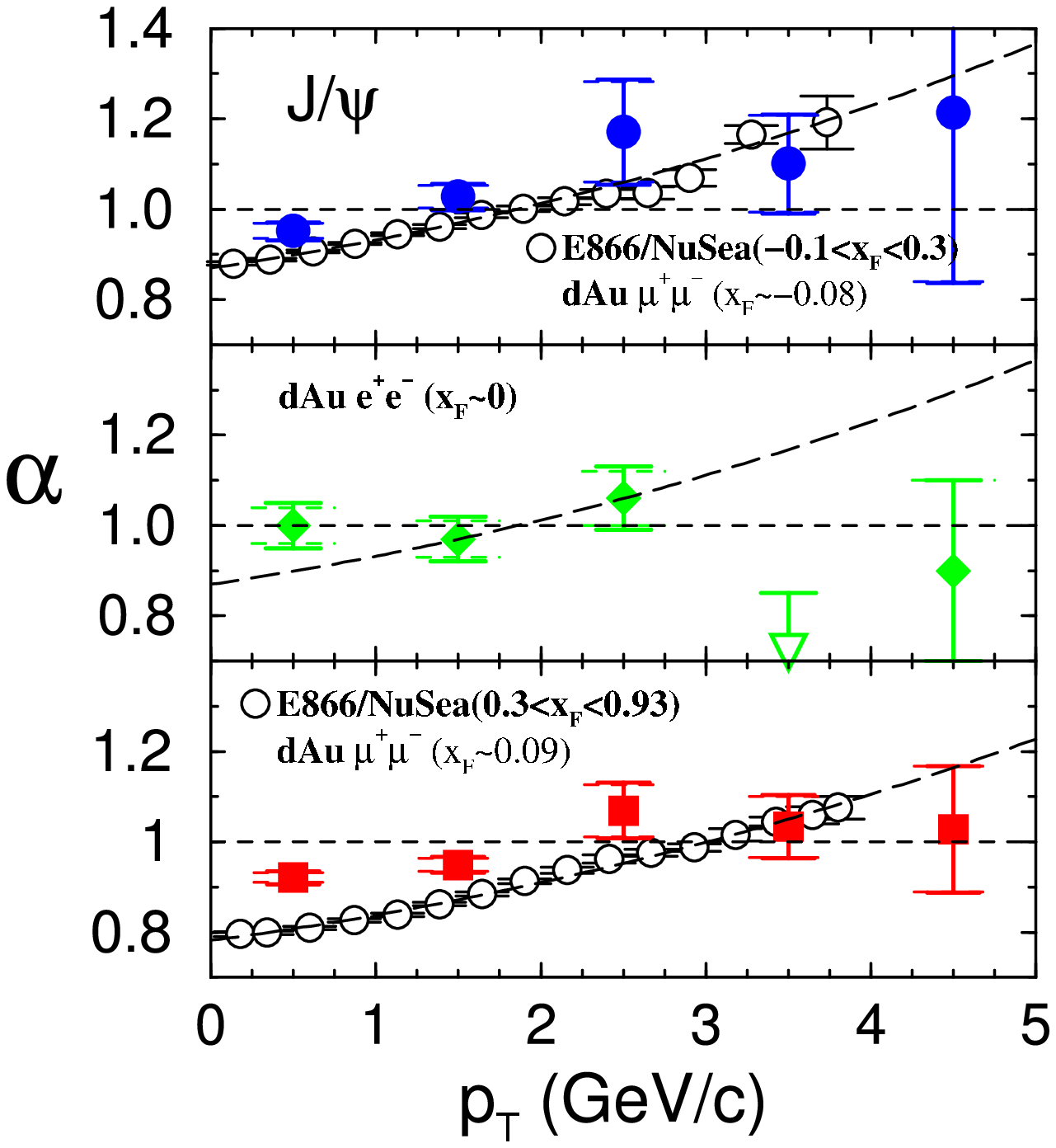}
\caption{\label{fig:alpha_pt2} $\alpha$ versus $p_T$ at different rapidity 
compared to E866 measurements.}
\label{fig:alphaPT}
\end{minipage}
\end{figure}

Figure \ref{fig:alphaPT} shows the nuclear modification factor 
$\alpha$ ($\sigma_{dAu} = \sigma_{pp}(2\times197)^{\alpha}$) versus $p_{T}$. An additional 0.02 overall uncertainty is not shown.  The dashed lines are simple fits to the E866 results\cite{E866}. 
One can see a similar tendency as shown in E866 results that the $\alpha$ 
increase towards larger $p_{T}$ in the forward and backward region. 
This is consistent with the fact that $<p_{T}^{2}>$ in 
$p+p$ collisions ($2.51\pm0.21GeV^{2}/c^{2}$) is much smaller than in $d+Au$ collisions (forward: $3.63\pm0.25GeV^{2}/c^{2}$, backward:$4.28\pm0.31GeV^{2}/c^{2}$) as expected from Cronin effect. 
One interesting observation is that the central arm measurements show no broadening.  The midrapidity $<p_{T}^{2}>$ in $p+p$ collision ($4.31\pm0.85GeV^{2}/c^{2}$) is
 larger by almost two standard deviations than that in the forward and 
backward rapidity.  It could either come from statistical fluctuation 
or indicate something interesting. The results from the high luminosity 2005 $p+p$ run should give a clear answer on this. 

Figure \ref{fig:alphaXbj}(a) plots the $\alpha$ against $x_{bj}$ in different 
collision energies. If shadowing or antishadowing is the 
dominant nuclear effect, one would expect that alpha scale with $x_{bj}$ since 
PDF is the function of $x_{bj}$ and $Q^{2}$ where $Q^{2} = m_{J/\psi}^{2}$ for 
$J/\psi$ production. This is in clear contradiction with the data. One possible 
explaination is that the coherence length increases with collision energy 
\cite{KopelivichCoherence}. The initial state energy loss dominate the nuclear 
effect at low energy (NA3) since the coherence length is short and the 
parton has large chance to lose energy before the hard scattering happens. At
RHIC energy shadowing is the dominant effect. We also plotted the alpha 
against $x_{F}$ as shown in figure \ref{fig:alphaXbj}(b) in different collision
energies. The result shows interesting scaling behavior. Some theoretical 
calculations try to explain the data with the Sudakov effect \cite{KopelivichSudakov}.

\begin{figure}[htb]
\begin{minipage}[t]{80mm}
\includegraphics[width=18pc, height=15pc]{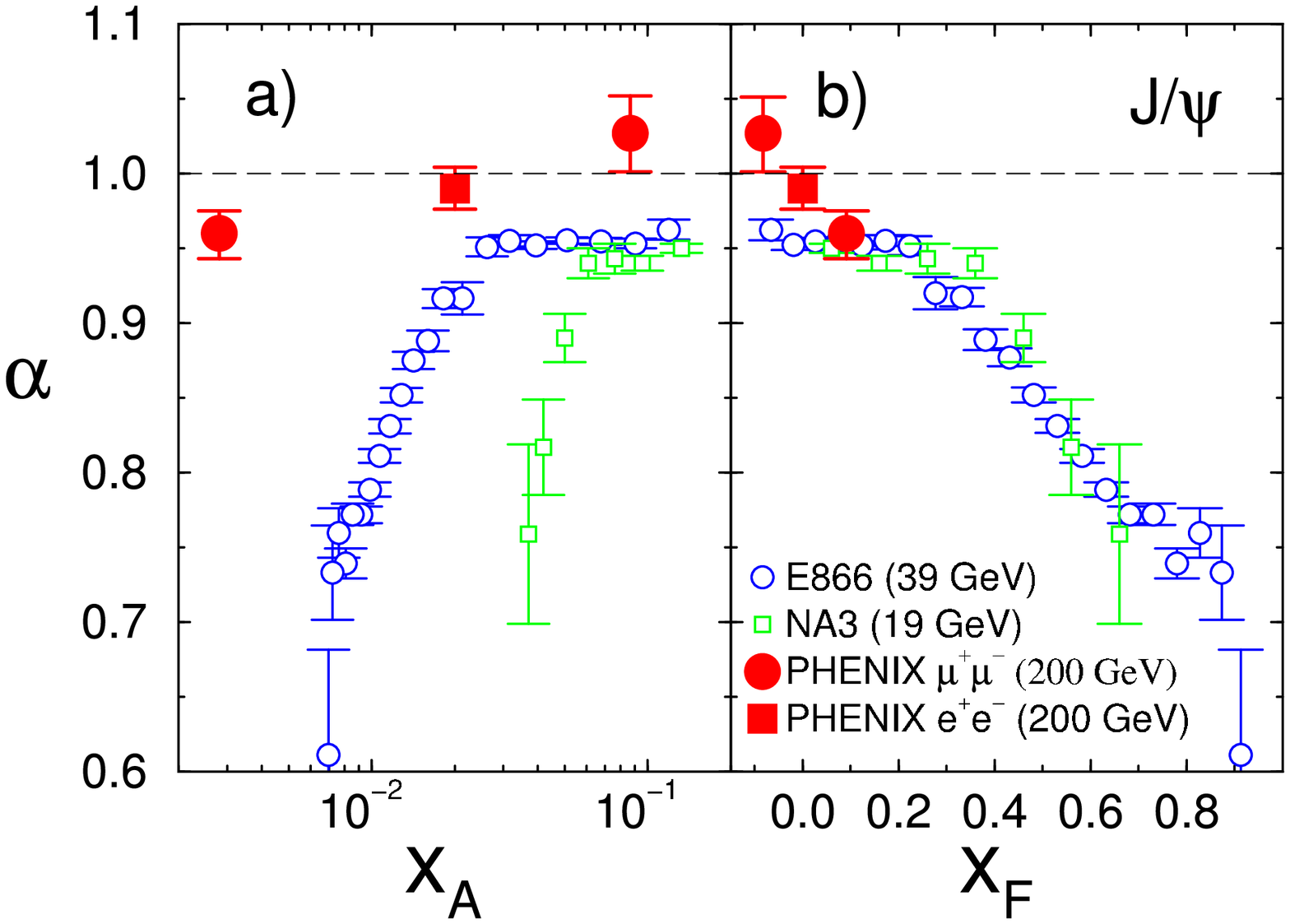}
\caption{ \label{fig:alphaXbj} $\alpha$ versus
(a) $x_A$ and (b) $x_F$ with present 200 GeV $J/\psi$ results compared to lower energy results \cite{E866}. An overall uncertainty of 0.02 in our $\alpha$ values is not shown.}
\label{fig:alphaXbj}
\end{minipage}
\hspace{\fill}
\begin{minipage}[t]{75mm}
\includegraphics[width=16pc, height=15pc]{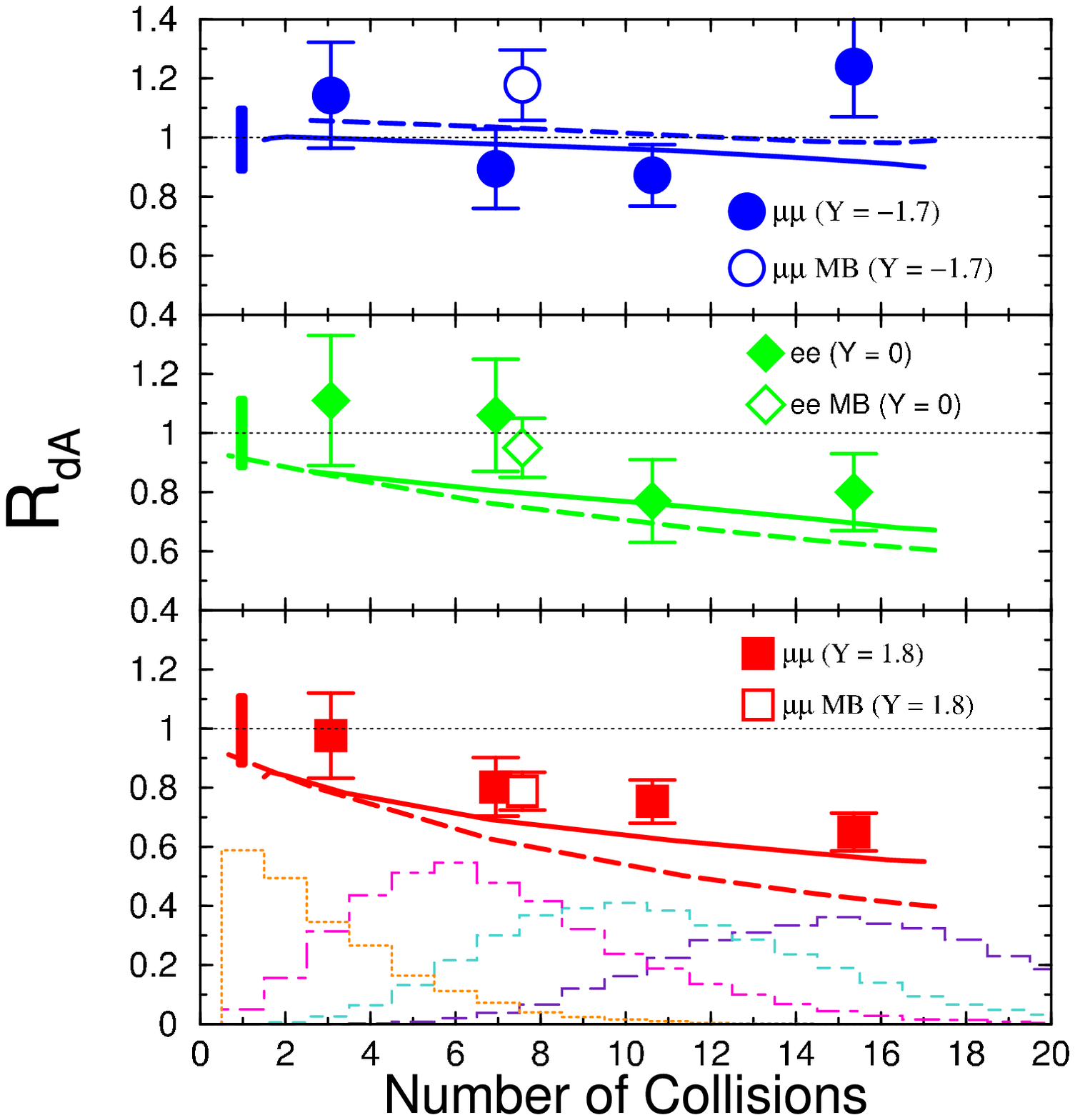}
\caption{\label{fig:rda_ncoll} $R_{dAu}$ versus centrality at three different rapidity regions, compared to calculations\cite{Vogt2004} including absorption and EKS (solid) or FGS (dashed) shadowing.}
\label{fig:rda_ncoll}
\end{minipage}
\end{figure}

Figure \ref{fig:rda_ncoll} shows the nuclear modification factor $R_{dAu}$  
versus centrality represented by the average number of collisions in each
 centrality bin at different rapidity regions.  The bars at the low end of each plot represent the scaling error in each rapidity region. An additional 12\% global error bar is not shown. The histograms at the bottom of the lower panel indicate the distribution of the number of collisions for each of the four centrality bins. One can see that in the small $x_{bj}$ region, the suppression 
become more significant towards more central collisions but the overall 
dependence is weak.  Similar but weaker centrality dependence is observed at midrapidity. 
No clear centrality dependency is found in the large $x_{bj}$.  Clearly the FGS
overpredicts the suppression in small $x_{bj}$ region.

One can deduce the baseline $R_{AuAu}$ at forward and backward rapidity via multiplying 
$R_{dAu}$ results in forward and backward rapidity. Similarly $R_{AuAu}$ at 
midrapidity can be deduced via multiplying $R_{dAu}$ at midrapidity by itself.
The $R_{AuAu}$ calculated this way as shown in figure \ref{fig:R_AuAu} does not
fully account for some nuclear effects like the final state comover suppression
but it tell us the amount of shadowing and absorption expected to be seen 
in $Au+Au$ collisions.  Theoretical calculations shown in figure \ref{fig:rapidity}(b) are also plotted after some rescaling to approximately describe the data. Same calculation is done to get the $R_{AuAu}$ prediction from these theory curves. One can see the interesting change of $R_{AuAu}$ shapes corresponding to different amount of shadowing and antishadowing effect.

\begin{figure}[htb]
\begin{minipage}[t]{75mm}
\includegraphics[width=16pc, height=15pc]{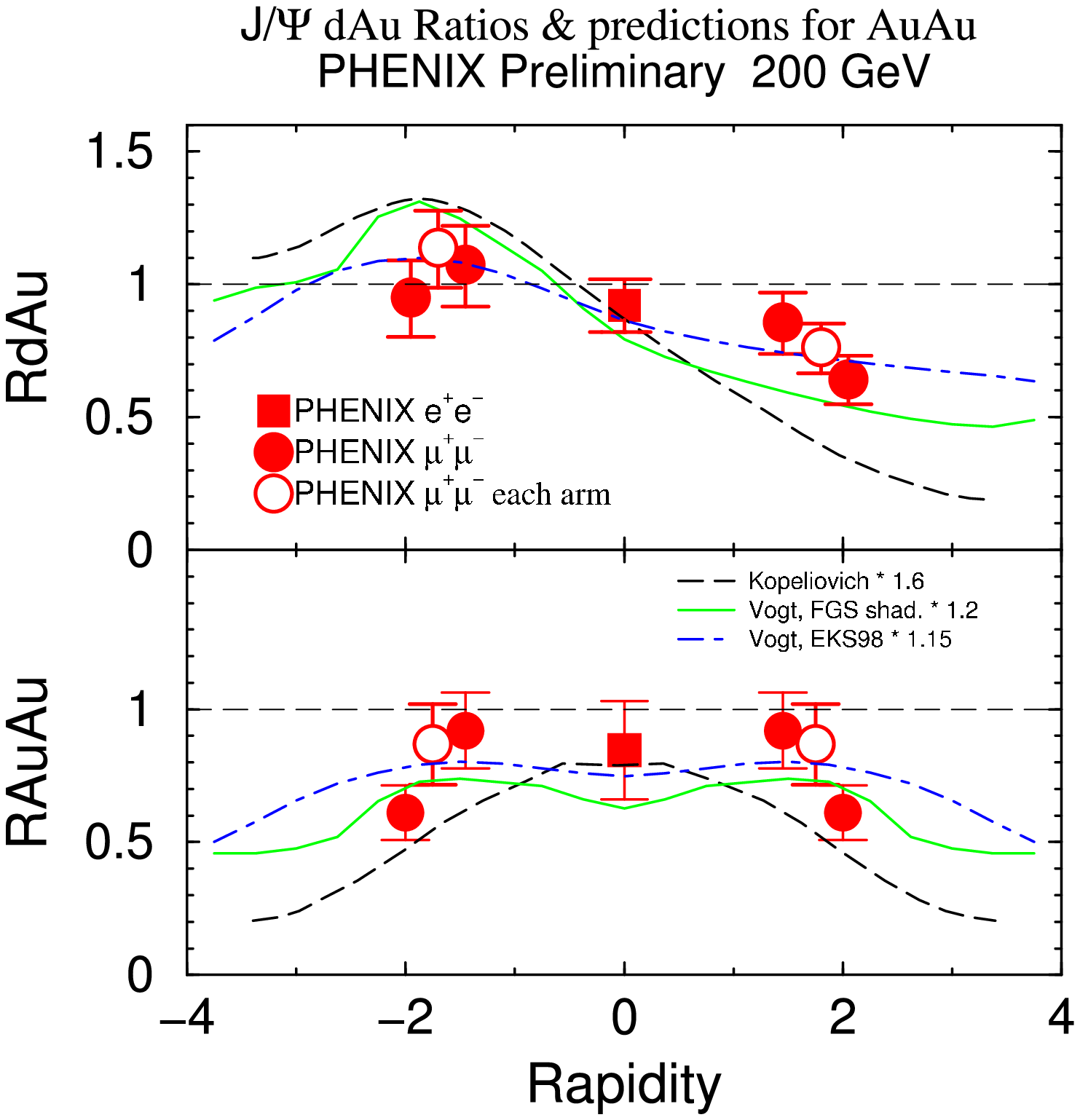}
\caption{ \label{fig:R_AuAu} $R_{AuAu}$ deduced from measured $R_{dAu}$ at different rapidity region.} 
\label{fig:R_AuAu}
\end{minipage}
\hspace{\fill}
\begin{minipage}[t]{75mm}
\includegraphics[width=16pc, height=14.5pc]{ups_y.ps}
\caption{\label{fig:upsilon} First $\Upsilon$ measurements at RHIC.} 
\label{fig:upsilon}
\end{minipage}
\end{figure}

\section{Summary and outlook}
In year 2003, PHENIX made the first measurement of cold nuclear effect on 
$J/\psi$ production in $\sqrt{s_{NN}}$ = 200~GeV $d+Au$ collisions at RHIC. The results show weakly increased $J/\psi$ suppression towards more central collision and smaller $x_{bj}$ region.  $p_{T}$ broadening is observed although the 
midrapidity results show no such effect possibly due to the statsitical 
fluctuation.  More luminosity of $d+Au$ collisions is needed 
to do detailed study on cold nuclear effect.  In year 2005, PHENIX 
accumulated much higher luminosity of $p+p$ collisions than that in year 2003 as
being illustrated by first measurements of $\Upsilon$ at RHIC shown in figure \ref{fig:upsilon}. This will significantly reduce the uncertainty of our $J/\psi$ measurements in cold nuclear medium.

\end{document}